# Advancement of Multifunctional support structure technologies (AMFSST)

R. John[(1)], G. Atxaga[(2)], H. J. Frerker[(3)], A. Newerla[(4)]

[(1)]HTS GmbH Am Glaswerk 6 D-01640 Coswig Germany, [(2)]INASMET Tecnalia Mikeletegi Pasealekua 2 Parque Tecnologico E-20009 Donostia-San Sebastian Spain, [(3)]OHB Systems AG Universitätsallee 27-29 D-28359 Bremen Germany, [(4)]European Space Agency ESTEC, Postbus 299 2200 AG Noordwijk Netherlands

*Abstract*- **The multifunctional support structure (MFSS) technology is promising a reduction of overall mass and packing volume for spacecraft (S/C) electronic components. This technology eliminates the electronic box chassis and the cabling between the boxes by integrating the electronics, thermal control and the structural support into one single element. The ultimate goal of the MFSS technology is to reduce size, weight, power consumption, cost and production time for future spacecraft components.**

**The paper focus on the main challenges and solutions related to the thermal management within the MFSS technology based on the selected charge regulator (CR) application.**

**Starting with the main set of thermal requirements for the CR the paper will include:**

- **Conceptual and detailed design based on high-conductivity carbon fibre CFRP,**
- **Description and results of the thermal material sample test program,**
- **Parameter and results for the performed first thermal simulation**

## I. INTRODUCTION

In recent years the trend in spacecraft design has been towards smaller, lighter and higher performance satellites with sophisticated payloads and instrumentation. As the launch vehicle size decreases, the satellites are weight and volume constrained. Therefore, the development of lighter electronic enclosures that meet the challenging requirements imposed by recent electronic trends is very interesting.

The multifunctional support structure (MFSS) or multi-functional support panel technology is a revolutionary concept of spacecraft architectures, mainly developed in the US at the end of the 90's. A demonstrator was flown and validated on the Deep Space 1 spacecraft as part of the New Millenium Programme (NMP) [1]. This technology eliminates the electronic box chassis and the cabling between the boxes by integrating the electronics, thermal control and the structure into a single element. The ultimate goal of the MFSS technology is to reduce size, weight, power, cost and production time for future spacecrafts.

A first initiative to develop such structure technologies for European spacecraft applications was undertaken in the past as part of the "Advanced Equipment Design " (AED) mini-project (ESTEC Contract No. 16725/02/NL/CK) [2] where a MFSS bread-board model was designed, manufactured and tested within the sub-activity "Multifunctional support panels for electronic equipment". The structure was based on lightweight composite panels of 500x280mm² dimensions.

The Multifunctional Structures Technology was applied and proved successfully. Rough estimates indicate that about 40% of weight reduction could be achieved by changing from aluminum to high conductive composite enclosures. Therefore, the successful development of a multifunctional structure, which integrates the required electro-magnetic performances, meets the thermal management challenges and maintains the structural properties will allow the European satellite manufactures to reduce the future launch cost.

MFSS combines different aspects from several engineering disciplines. Therefore, it is not only a technological task to integrate the desired functions into a part but also a question of effective concurrent engineering related to team integration and system design approach.

The result of the AED project was that thermal and mechanical requirements could be achieved by means of the use of high thermal conductivity carbon fibres combined with high strength carbon fibres (i.e. K1100 with M40J). However, radiation protection, electro-magnetic compatibility (EMC) and electro-magnetic interference (EMI) shielding of these enclosures was an open point.

Therefore, the current project – "Advancement of Multifunctional support structure technologies" (AMFSST) (ESTEC Contract No. 19944/06/NL/PM) will be focusing on the following fields:

- CFRP manufacturability by using high thermal conductive fibres,
- EMI and EMC shielding of composite housings,
- Radiation shielding,
- Electrical bonding.

The main scope of the project is to validate a MFSS prototype for a relevant spacecraft subsystem. Therefore, design, analysis, manufacturing and test of the prototype will be performed.

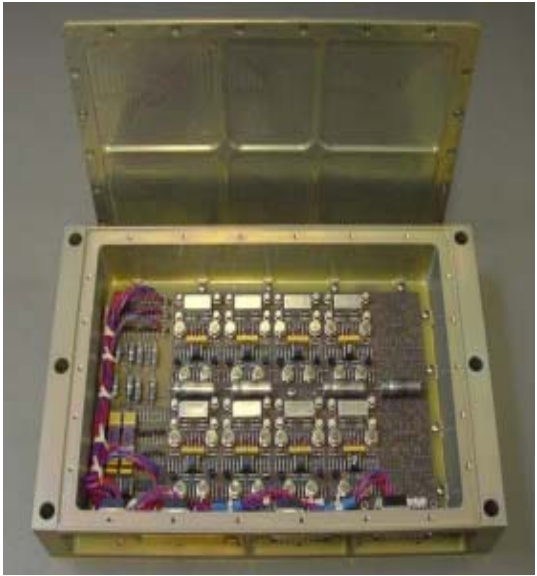

Fig. 1. CR in conventional aluminum housing (photo: OHB Systems AG).

As a reference application a charge regulator (CR) unit as depicted in Fig. 1 was selected. The charge regulator as part of the Power Control Unit (PCU) controls the main satellite power bus respective by the battery voltage and the battery charge current by shunting to ground solar panel strings.

The activities reported in this paper have been executed within the AMFSST.

## II. MAIN REQUIREMENTS AND SPECIFICATIONS

Each requirement contained in this section will be verified by a suitable method. These requirements base on the performed selection of the CR as reference application.

Though this paper focuses on the thermal aspect of the multifunctional structure the main structural and electronic requirements will be stated too, because there are of course relations between all three functionalities of the structure.

- REQ-MFSS-MAS-010: *Mechanical stiffness* for the MFSS shall be higher than 100Hz
- REQ-MFSS-MAS-012: The CR Design within the MFSS shall cover a *volume* not exceeding: Length: 296 mm, Width: 189 mm, Height: 53 mm The volume should be minimized as much as possible
- REQ-MFSS-MAS-015: The MFSS *thermal control* system shall be able to radiate a maximum heat dissipation with α < 0.3 and ε > 0.7
- REQ-MFSS-ENV-001: The CR (EM and FM / PFM) shall be able to sustain the following e*nvironmental temperatures* without degradation of performance: Prototype: -40°C to +70°C operational, EM: 0°C to +50°C operational 0°C to +70°C non operational (storage), FM/PFM: -40°C to +70°C operational -55°C to +125°C non operational (storage)
- REQ-MFSS-IFR-010: The CR based shall be realized as structural *sandwich panel* fabricated using high thermal conductivity (Hi-K) composite skins.
- REQ-MFSS-IFR-014: The *mass* of the CR shall be less than 1980g.
- REQ-MFSS-IFR-008: The power consumption of the CR shall be in the range of 1.1W.

## III. MATERIAL SELECTION

Composite materials are central in space applications due to the weight savings that could result from using low density polymer matrix composites made from high modulus and high strength fibres. The main advantages of composites include high specific strength (strength per unit weight) and stiffness, corrosion and fatigue resistance, tailorable thermal conductivities, controlled thermal expansion and the ability to be processed into complex shapes. Advantage and challenge at the same time is the possibility to design all the thermo-mechanical properties as orthotropic ones.

Pitch based high thermal conductivity carbon fibres are promising materials, that possess superior specific stiffness and thermal conductivity than PAN-based fibres.

The main goals for the development of carbon fibre for space applications are

- High stiffness and strength,
- High thermal conductivity,
- Good handling ability, and
- Acceptable cost performance.

A summary table with a selection of the most important high conductivity fibres is presented in Table 1.

Based on the requirements and the experiences from former projects the carbon fibre K13D2U from Mitsubishi was selected for application in this project. It seems to be a good compromise between thermal conductivity and handling ability.

Table 1
High thermal conductivity carbon fibres (continuous fibres)

| Type | Name | Manu-facturer | Tensile Modulus [GPa] | Thermal conductivity [W/(m K)] | Density [g/cm³] |
|------|------|------|------|------|------|
| Pan | M40J | Toray | 372 | ~40 | 1.77 |
| Pitch | K13B2U | Mitsubishi | 827 | 260 | 2.16 |
| | K13C2U | | 900 | 620 | 2.2 |
| | K13D2U | | 930 | 800 | 2.2 |
| | YSH-70A | NGF | 724 | 250 | 2.16 |
| | YS-90A | | 896 | 430 | 2.19 |
| | YS-95A | | 924 | 600 | 2.19 |
| | P-120S | Cytec (BP Amoco) | 828 | 600 | 2.16 |
| | K800 | | 896 | 800 | 2.2 |
| | K1100 | | 931 | 1000 | 2.2 |

There are several considerations that must be taken into account when selecting the polymeric resin for space application. Summarizing, the essential parameters are

- Outgassing under vacuum,
- Micro-cracking and
- Resistance against atomic oxygen.

Candidate resin types for such application are

- Toughened epoxy resins,
- Cyanate ester resins, and
- Bismaleimide resins.

*Epoxy resins* have been the baseline for space composites manufacturing because they offer a unique combination of properties that are unattainable with other thermosetting materials. Nevertheless, first generation epoxies for space experienced several drawbacks: the high humidity absorption in ground and subsequent desorption in orbit, dimensional instability and micro-cracking susceptibility.

*Cyanate ester* are a family of high temperature thermosetting resins, which have also other desirable characteristics that may justify their higher cost in particular for space applications. They possess a unique balance of properties and are particularly notable for their low dielectric constant and dielectric loss (radome applications), low moisture absorption and low shrinkage (dimensionally stable structures) and low outgassing characteristics.

*Bismaleimide resins* are a relatively young class of thermosetting polymers that are gaining acceptance by their excellent physical property retention at elevated temperatures and in wet environments, almost constant electrical properties over a wide range of temperatures and non flammability properties.

In the world there are only a few manufacturers of prepregs made from high conductivity carbon fibres, because of the high modulus and low ultimate strain of those fibres.

Within this project we selected the cyanate resin EX1515 and ordered the related prepregs from the company Bryte in the US.

So the final material selection is a compromise between material properties of fibres as well as matrix material, availability and costs.

A sandwich panel is the first choice for the basic structure to withstand the mechanical loads. The face sheets of the panel will be made of the CFRP material (K13D2U/EX1515) and the core will be an aluminum honeycomb. The Hexcel core 1/8-5052-S-0.0007-25 was selected for this application.

## IV.  MATERIAL TEST RESULTS

Several technological and material samples were made to ensure a good and constant manufacturing quality and to investigate the material as well as part properties related to the applied manufacturing technology.

Table 2
Thermal conductivity (tc) test plan

| Test No. | Test characterization | Kind of sample | Test method | Sample size [mm³] |
|---|---|---|---|---|
| 1a | In-plane tc longitudinal | Unidirectional CFRP (face sheet) | HFM | 150x20x0.75 |
| 1b | In-plane tc transversal | | HFM | |
| 2a | Through thickness thermal conductivity | Unidirectional CFRP (face sheet) | LFM | 10x10x0.75 |
| 2b | | | HFM | 30x30x0.75 |
| 3 | Through thickness thermal conductivity | Sandwich with quasi-isotropic face sheets | HFM | 30x30x26.5 |
| 4a | Through thickness thermal conductivity | Flex circuit bonded on CFRP face sheet | HFM | 30x30x1 |
| 4b | | Flex circuit bonded on sandwich | HFM | 30x30x27 |

The test campaign comprises mechanical, electrical and thermal tests. The thermal properties were measured by laser flash method (LFM) and an adapted heat flow meter (HFM) test stand for parts.

The thermal test plan is characterized in Table 2.

First test results were obtained for tests 1a and 2b and will be presented in the following. The basic test setup for the HFM method is sketched in Fig. 2. The results of the measurement (temperatures at the sample and the test stand, heat power input) are analyzed with a corresponding FE-simulation (optimization of sample temperatures) to calculate the heat loss over the surrounding insulation material. The measured fibre volume fraction for the applied manufacturing technology was 0.53 and the resulting thermal conductivity (test 1a) was 420.5 W/(m K).

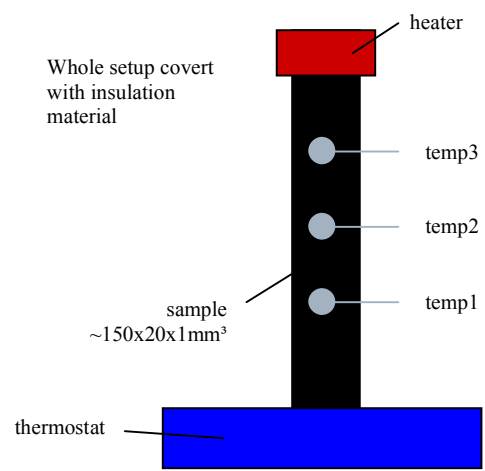

Fig. 2. Sketch of HFM test stand principle.

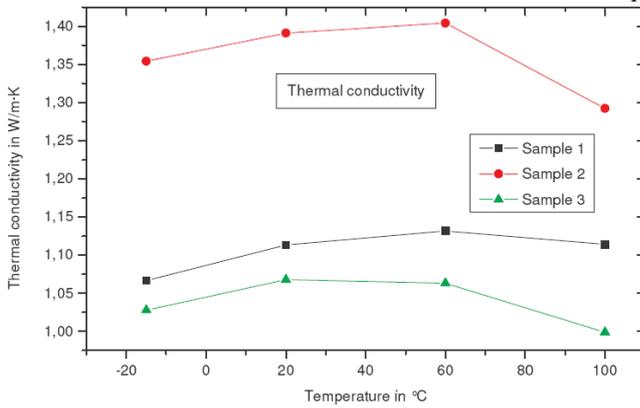

Fig. 3. Results for test 2a – through thickness thermal conductivity.

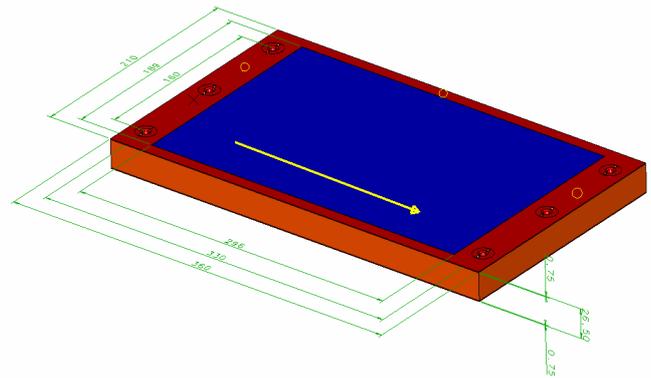

Fig. 4. Conceptual design.

For determination of the thermal conductivity with the LFM it is necessary to determine the thermal diffusivity a, specific heat capacity $c_p$ and the mass density ρ. The results for this measurement are plotted in Fig.3.

From the measured values it could be seen, that the thermal conductivity of an unidirectional CFRP part will be anisotropic: 420W/(m K) in longitudinal and about 1.2 W/(m K) in transversal direction.

The measured values will be used in the subsequent first analysis of the thermal prototype behavior.

## V. CONCEPTUAL DESIGN

The basic thermal functionality of the redesigned CR panel is

- to distribute the heat at the surface on which the CR PCB is placed
- to conduct the heat through the thickness to the opposite face sheet and
- to radiate the heat from this face sheet to the external environment.

The mass of the PCB is 0.8 kg, the box mass including harness is about 1.5 kg. Keeping the PCB-mass constant, the structural mass must not exceed 1.98 kg minus 0.8 kg, i.e. 1.18 kg. Assuming a sandwich panel of area dimensions as given for the CR, the mass will be far below the required 1.18 kg.

For given dimensions the lowest eigenfrequency of the multifunctional panel structure depends on the sandwich height, the face sheet thickness and Youngs modulus, and on the mass. The face sheet thickness is also controlled by the thermal conductivity. A very thin sheet would avoid a heat flux along the panel, even if the conductivity of the panel is very high.

The specified random vibration environment results to a load of about 15g RMS. In order to include almost all load peaks (with 99.7% probability) a factor 3 is commonly applied. With that we have a design load of 45g which is in line with the BAMFORD mass/acceleration curve for a structure between 1 and 2 kg.

For thermal reasons, the fixation to the structure should allow best thermal contact. This is for the given dimensions met by 6 bolt fixations.

It could be presumed, that not the mechanical load is the driving factor for dimensioning.

The former CR-box has overall dimensions of 296x189x53 mm³. The box includes PCB, sockets, harness and the mechanical structure. For integration of the PCB-content, the overall dimensions of the box will be necessary for the MFSS, too. So the panel must be actually larger than the CR-items – 360x210x26.5mm³. In the design it is assumed, that at least the inserts have to be outside of the area where the electrical parts will be integrated.

The selected material for the face sheets is the fibre K13D2U, which shows a good compromise between workability and fibre conductivity. The fibre conductivity of 800 W/(m K) in fibre direction will give an quasi-isotropic in-plane part conductivity of about 200 W/(m K), i.e. in the range of aluminum (according to simple mixing formulas for the thermal conductivity of laminates).

Assuming 45 g as the design load then the force onto the 6 fixation points is in the range of 500 N, i.e. per fixation point the load is about 100 N. This value is uncritical for both, bolt and basic material.

Fig. 4 shows the conceptual design of the engineering model. It consists of:

- the sandwich panel [brown] (CFRP face sheets – 0.75mm thickness – with aluminum honeycomb – 25mm thickness – and 6 aluminum inserts for fixation), and

- the reserved area for CR PCB with connectors and harness [blue]. The yellow arrow marks the 0° direction for the fibre layup.

A quasi-isotropic stacking sequence for the face sheet laminate layup will be well suited for distribution of the heat load within the face sheet. For both face sheets the same stacking sequence of 8 layers is used: [0°,90°,45°,-45°]symmetric.

## VI. THERMAL SIMULATION

A steady state thermal simulation with Ansys was performed to get a first impression of the thermal performance of the redesigned CR-panel.

Modern FE-codes (e.g. Ansys, I-deas TMG) do not provide layered element formulations for anisotropic thermal conductivity. So the face sheets of the panel could be modeled in two ways:

a) by analytical pre-calculation of the anisotropic material parameters based on simple mixing rules (analogous to the classical laminate theory for mechanical properties) and application of the gained parameters to a one element thick volume, or

b) by modeling the real stacking sequence of the laminate (8 volumes other the face sheet thickness) and assignment of the corresponding layer material properties according to their orientation.

The overall behavior of the model should be independent of the modeling strategy but some result details may differ.

The main model parameter are (please see Fig. 5):

- Geometry: height 210mm, length 360mm, thickness: core 25mm, face sheets 0.75mm, heat power concentration areas: count 10 size 20x20mm²

- Boundary conditions: load: 1.1W equally distributed to the 10 heat power concentration areas, heat transfer modes: conduction within the panel and radiation from panel back side to the environment (ambient temperature -180°C)

- Material parameters: thermal conductivities in [mW/(mm K)] Core: kxx=1.16,kyy=0.77,kzz=2.07 [3]; face sheets: model a) kxx=kyy=213.5, kzz=0.4 model b) 8 layers [0,90,45,-45] kxx=424.9, kyy=kzz=0.4, emissivity of face sheets 0.8

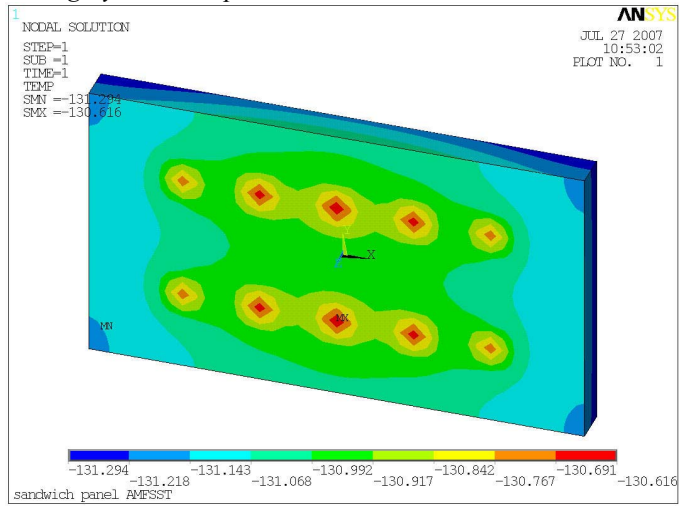

Fig. 6. Temperature distribution for model a).

The FE-model consists of the sandwich panel with anisotropic CFRP face sheets and anisotropic aluminum honeycomb. Both parts are meshed with solid70 elements. The volume is partitioned to get 2*5=10 heat power concentration areas each with a size of 20x20mm². On the opposite face sheet the solid70 elements are covered with surf152 elements for modeling the heat transfer to the environment by radiation.

Fig. 6 and Fig. 7 show the results for model a) (one element over face sheet thickness) and b) (8 elements over face sheet thickness). It could be seen, that the temperature level is the same for both models but the distribution is different due to the influence of the staking sequence. Mainly the 45° layers seem to disturb the model symmetry. It is also seen, that the temperature level is too low.

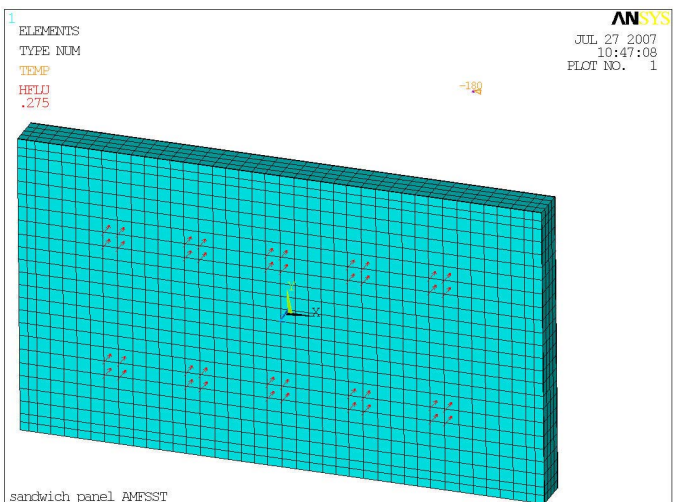

Fig. 5. Applied FE-model with boundary conditions.

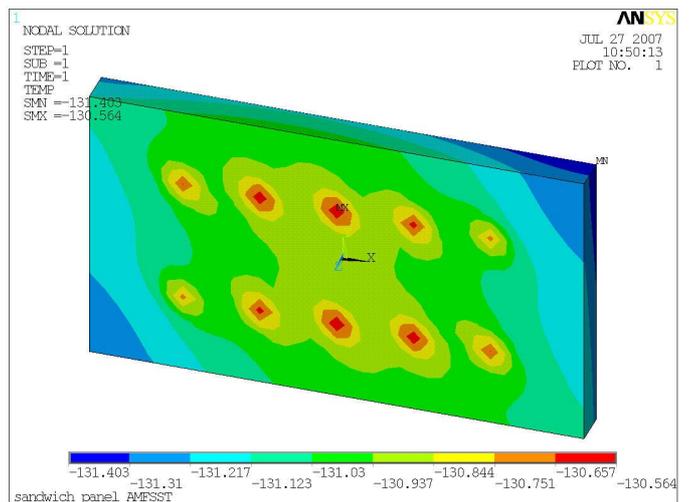

Fig. 7. Temperature distribution for model b).

There are two possibilities to raise the temperature level:

1) using the panel radiator also for power dissipation from adjacent electronic components or
2) reduction of the radiation surface for instance by application multi layer insulation (MLI).

For a reduced radiation area to a quarter of the whole panel the mean temperature raise to ~-26°C as seen in Fig. 8.

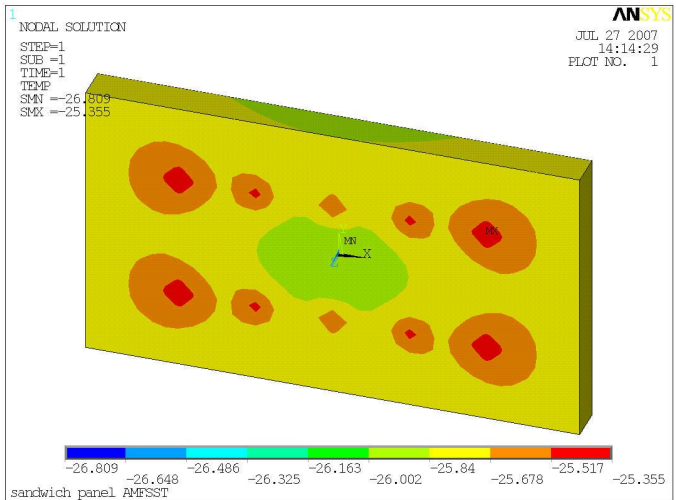

Fig. 8. Temperature distribution for model b) with reduced radiation area.

## VII. CONCLUSIONS

It was shown, that from thermal management point of view a redesign of the classical aluminum housing for a space application CR into a multifunctional sandwich panel with face sheets made from high conductivity carbon fibres is feasible.

In the next project steps we will manufacture and test a prototype of such CR to get among others measurement data for comparison with the thermal simulation results.

The FE-modeling of a layered structure with anisotropic thermal conductivity isn't a today's standard task. In this field some further investigations will be necessary. Especially we will perform a detailed comparison between the gained test and simulation data.

## ACKNOWLEDGMENT

The activities reported in this paper have been executed within the ESA General Support Technology Programme (GSTP) contract 19944/06/NL/PM "Advancement of Multifunctional support structure technologies".